
\magnification=\magstep1
\font\B=cmbx10 scaled\magstep1
\font\cs=cmcsc10
\baselineskip=21pt

{}~~~~
\vskip.5truein

\centerline{\B Branching Processes and Multi-Particle Production}
\bigskip\bigskip

\centerline{{\cs S. G. Matinyan}\footnote{*}{On leave from Yerevan Physics
Institute, 375036, Yerevan, Armenia}}

\centerline{Department of Physics, Duke University}

\centerline{Durham, North Carolina 27708-0305}
\medskip

\centerline{\cs E. B. Prokhorenko}

\centerline{Theory Division, Yerevan Physics Institute}

\centerline{375036, Yerevan, Armenia}
\vskip.5truein

\centerline{\bf Abstract}
\medskip

{\narrower
The general theory of the branching processes is used for establishing
the relation between the parameters $k$ and $\bar n$ of the negative
binomial distribution.  This relation gives the possibility to
describe the overall data on multiplicity distributions in $pp
(p\bar p)$-collisions for energies up to 900 GeV and to make several
interesting predictions for higher energies.  This general approach
is free from ambiguities associated with the extrapolation of the
parameter $k$ to unity.
\bigskip\bigskip}

\noindent {\bf 1. Introduction}
\medskip

Theoretical description of the multiple production, based on so-called
soft processes, today is beyond the limits of QCD, and the natural
approach there is to look for empirical relations.

The most popular in this field was the KNO scaling for multiplicity
distributions which was fulfilled very well for hadronic collisions up to
ISR energies and for $e^+e^-$-annihilation.  The evident violation of
the KNO scaling at energies of the CERN collider [1] attracted much
attention to the negative binomial distribution (NBD) which describes
fairly well the overall features of the data on the multiplicity
distribution (MD) of hadrons in different processes $(pp(\overline{p}p)$,
$e^+e^-,\nu p, AA \ldots)$, in different ranges of rapidity and in a wide
interval of energies [2,3].  It is especially relevant to the
$pp(\bar pp)$ interaction.

Taking into account the special role of NBD in describing the multiplicity
distribution at high energy, it seems to be important to consider NBD
on the basis of general assumptions about the character of the process
of particle production in hadronic collisions without detailed
specification of the dynamics.  In reference [4] it was proposed as a
basis for NBD to consider the multiple production as a random stationary
branching process which is a rather general probabilistic model for the
processes of the multiplication and transformation of the active particles.
In this approach the transformation of each particle is independent of
the history of the process and of the transformation of other particles,
obeying the general probabilistic laws of Markov processes.  The same
refers to the fate of the generation of each particle.

The branching processes may find their realization in terms of the
quark-gluonic cascades, corresponding to the microscopic description of
the nonequilibrial evolution of the partonic system, e.g., in the
rapidity space [5,6].  It is important to stress that for us there is no
need to know the details of the dynamical laws governing these cascades.

It was realized that the system of produced hadrons may be considered as
a result of the contribution from coherent and chaotic components (so
called two-component model) and it is known that in $pp(\bar pp)$
collisions at high energy the chaotic component [7-10] dominates,
which described by NBD, whereas in $e^+e^-$-annihilation the coherent
(Poisson) part is essential.

The observation of dynamical chaos in the dynamics of nonabelian gauge
fields (see e.g. [11]) raises the question about the role and
origin of the chaotic component in hadronic collisions.

There exists an interesting practical observation [7,9], that in
distinction to $e^+e^-$-annihilation where the addition of the small
($\approx$ 10-20\%) chaotic (noise) amplitude essentially changes
the multiplicity distribution, in $pp(\bar pp)$-collisions the
addition of a small coherent component to the NBD does not change the
shape of the distribution significantly.

Taking into consideration the above mentioned arguments and remarks we
here consider the NBD as adequate for the description of the multiplicity
distribution in $pp(\bar pp)$ collisions at high energy.\footnote{$^1$}
{In principle on the basis of the quantum optics it is possible to
generalize NBD to take into account the chaotic as well as the coherent
components [7,12].}  The NBD
$$P_n^{(k)} = {\Gamma(n+k)\over \Gamma(n+1)\Gamma (k)}\;\;
{\left({\bar n\over k}\right)^n \over \left( 1+{\bar n\over k}\right)^{n+k}}
\eqno(1)$$
has two parameters $\bar n$ and $k$.  $\bar n$ is the average
multiplicity.  As for $k$, initially it was associated with the number
of chaotically emiting cells.  After the UA5 experiments [1,10,14] it
is clear that such a meaning of $k$ in general is not necessary,
because $Sp\bar pS$ collider data yield the empirical relation
$${1\over k} = a+b\ln\sqrt{s} \quad (a\approx -0.1,\; b\approx 0.06)
\eqno(2)$$
which is valid up to energy $\sqrt{s}$ = 900 GeV, not showing a
tendency for saturation.  So, at such energies the KNO-like scaling
continues to be violated, which is more clearly expressed in the observed
strong rise of the moments
$$C_q = {\langle n^q\rangle \over \langle n\rangle^q}\eqno(3)$$
with $\sqrt{s}$ [14].

Of course, one cannot extrapolate the concrete form of the empirical
relation (1) to higher energy since this would lead to contradiction.
For instance, from (1) it would follow that the peak of the distribution
would be at $n=0$ at very high energy when $k=1$.  This means that
saturation of $k$ must take place at ultrahigh energies at a value
larger than unity (see also [15]).  It indicates the necessity to
establish the relation between $k$ and $\sqrt{s}$ (or, at least,
between $k$ and $\bar n$) based on general theoretical considerations.
We propose that such a basis could be a general theory of branching
processes.  As mentioned such an approach was developed in [4] where
the idea of the {\it stationarity} of the branching process was used
for establishing the relation between $k$ and $\bar n$.  The result
$${1\over k} = a + b \ln {\bar n\over k} \qquad (a \approx 0.12,\;
b\approx 0.08) \eqno(4)$$
gives the unconfined though weaker rise of ${1\over k}$ with $\sqrt{s}$
leading to the above difficulty associated with the extropolation of $k$
to unity.  Unfortunately, in deriving (4) the authors of [4] incorrectly
used the conditions for stationarity of the branching process.  In the paper
[16] the condition for stationarity is used correctly though the
authors missed the most interesting ansatz, in our opinion, of the
relation between $k$ and $\bar n$.  The resulting dependence of $k$ on
$\bar n$,
$$k = A\left( {\bar n\over k}\right)^B (A \simeq 11,\; B\approx
-0.5)\eqno(5)$$
again did not avoid the problem resulting from the extrapolation of $k$
to unity.
\bigskip

\noindent {\bf 2. Relation between $k$ and $\bar n$}
\medskip

Thus, we consider NBD as a result of the stationary branching process
with one sort of multiplied particles (pions) and continuous evolution
parameter $t$.  The generating function $F$ for such a process satisfies
the reverse Kolmogorov differential equation [17]:
$${dF\over dt} = f(F,t). \eqno(6)$$
For the generating function of the NBD
$$F(x,t) = \sum_n P_n^{(k)} x^n = \left[ 1 + {\bar n\over k}
(1-x)\right]^{-k}\eqno(7)$$
$f(F,t)$ equals
$$-F \ln F {\dot k\over k} + F(1-F^{1/k}) k {\dot m\over m}, \eqno(8)$$
where $m = \bar n/k$ and $\dot k = dk/dt$, etc.

For a stationary branching process $f(F,t)$ is factorized,  $f(F,t)
= \varphi (F) \psi (t)$.  Evidently the condition
$${\dot k\over k} = \hbox{const.}\; k {\dot m\over m}\eqno(9)$$
[4] which leads to the relation (4) does not give such a factorization.
For $F\approx 1$ factorization takes place [16] and this approximation
is also good for $k$ close to unity.  More adequate here is a parameter
$$\delta = {\ln F\over k}, \eqno(10)$$
which is small at $F \approx 1$ and not too small $k$.  Expanding
$1-F^{1/k}$ up to $\delta^2$ in (8), it is easy to obtain the solution
of the resulting differential equation which is a necessary and sufficient
condition for factorization:
$$k = {a\bar n\over \bar n-b}, \eqno(11) $$
where $a$ and $b$ are the integration constants which one must find
from comparison with experimental data.  Thus it is possible to state
that NBD with relation (11) between its parameters $k$ and $\bar n$ is
the consequence of a stationary branching process.

The function $k(\bar n)$  is very simple.  At $ab>0$ $k$ is decreasing
from $a$ to $-\infty$ and from $+\infty$ to $a$.  But {\it experimentally}
(at least for $10 < \sqrt{s} <$ 900 GeV) $k$ is decreasing with $\sqrt{s}$
[1,14,10], so physically interesting is a case $ab > 0$.  But at the same
time the case $a<0$, $b<0$ is also unphysical, because it corresponds to
$\bar n < 0$, or to $k<0$.  By the same reason, if we do not want to have
negative $k$, we must discard the lower branch of (5) with $ab>0$
corresponding to decreasing $k$ in $(a, -\infty)$ interval.\footnote{$^2$}
{In this connection there is an interesting observation in [18] that the
multiplicity data for small $(\sqrt{s}<10\hbox{GeV})$ energies is
possible to describe well with negative $k$.  From this point of view
it may be said that two branches of the hyperbola (5) naturally divide
the large energy region $(\bar n>b>0)$ from small energies $(\bar n< b)$.
{}From our fit (see below) $b\approx 7$, so it means that we must not
consider energies below $\sqrt{s} \approx$ 14 GeV.}

Thus this simple analysis has shown that in the region of high energy
the relation between $k$ and $\bar n$ is given by (5) with positive $a,b$,
and it is necessary to confine to the branch of the hyperbola (5) in
the first quadrant.
\bigskip

\noindent {\bf 3. Consequences and predictions of the model}
\medskip

The relation (5) between $k$ and $\bar n$, in spite of its simplicity,
is rather rich in content.  Let us stress once more that this relation
must be only used for $\bar n > b \approx 7$, i.e. for $\sqrt{s} > 14$
GeV; smaller energies should not be considered here.  Our model ensures
that $k > a> 0$ (from the fit follows $a$ = 3.06), implying that the
limit $k=1$ never is achieved.  Thus there does not exist the difficulty
associated with $k=1$ at very high energy that is characteristic of
some other  ansatze [1,4,16].

{}From (5) it follows that asymptotically, when $k$ goes to the
saturation, KNO scaling is restored.  The asymptotic distribution function
at $\sqrt{s} >>$ 14 GeV and $\bar n \gg k = a\approx 3.06$ has a form of
$\Gamma$-distribution:
$$\psi (z) \equiv \bar n P_n^{(k)} \approx {k^k z^{k-1} e^{-kz}\over
\Gamma(k)} = 14.49 z^{2.06} e^{-3.06z} \quad \left( z =
{n\over \bar n}\right) \eqno(12)$$

Our model gives rather clear predictions for $C_q$-moments.  In
particular, Wroblewski's relation here takes place well at high
energies:
$${D_2\over\bar n} \equiv {\sqrt{\overline{n^2}-\bar n^2}\over\bar n} =
(C_2-1)^{{1\over 2}} = \left( {1\over k}+{1\over\bar
n}\right)^{{1\over 2}} \approx 0.57 \left( 1-{1.92\over\bar
n}\right). \eqno(13)$$

The second order correlation $g^{(2)} = {\overline{n(n-1)}\over\bar n^2}$
which is increasing slowly and asymptotically equals 1.33 indicates not
the presence of a coherent component as sometimes stated but just the
fact that $k$ is always larger than unity.  All high order moments $C_q$
are rising and saturate asymptotically, as is easy to understand from the
relation $(q>2)$
$$C_q = 1 + \sum_{m=0}^{q-2} P_m^{(q)} \left({1\over k}\right)
(C_2-1)^{q-m-1} \eqno(14) $$
where $P_m^q\left({1\over k}\right)$ are polynomials of order $m$ with
positive coefficients $(P_0^{(q)}=1)$ and ${1\over k} = {1\over a} -
{b\over a} {1\over \bar n}$ is increasing.  Asymptotically we have:
$$ C_q  \approx {\Gamma(k+q)\over k^{q-1}\Gamma(k+1)} \eqno(15)$$
i.e. (up to $1/\bar n^2$):
$$\eqalignno{ C_2 &\approx 1.33- {1.27\over \bar n} \cr
C_3 &\approx 2.21 - {5.82\over \bar n} \cr
C_4 &\approx 4.39 - {21.3\over\bar n} \cr
C_5 &\approx 10.19 - {76.11\over\bar n}. &(16)\cr}$$

The scaled peaks of the multiplicity distributions is moving to the left
toward its asymptotic value:
$$z_{\rm peak} (\sqrt{s})= {n_{\rm peak}\over \bar n} = 1-{1\over a} +
{b\over a} {1\over\bar n} = 0.67 + {2.27\over\bar n}. \eqno(17)$$

Finally, before going into the comparison with experimental data, let
us make one comment.  By no means do we consider the limiting value of
$k=a$ as an indication that corresponds to asymptotic value of the number
of clusters, fireballs, minijets etc. in the multiple
production.\footnote{$^3$}{Note that some experiments (see [20]) are
indicating that at sufficiently high energies ($E_L \approx$ 400 GeV)
the number of the clusters produced in $pp$-interactions is $4.2\pm 1.7$.
If we continue such an interpretation of $k$, then $k^{-1}$ may be
considered as the ratio of the probability for two particles to be
emitted from one cluster to the probability of emission of these particles
by two different clusters [6].  So the asymptotic ``aggregation''
degree is $a^{-1} \approx 0.33$.}

Let us stress only that the often used value of $k=1$ is
meaningless.\footnote{$^4$}{The $k=1$ in our model corresponds to the
negative $\bar n$ (lower branch of (5) from $a$ to $-\infty$).  Notice
that $k > a \simeq 3$ shows that our expansion parameter of $\delta =
k^{-1} \ln F$ is adequate and selfconsistent.} In particular, in
connection with this value of $k$ in [16] it was made very strong and
unusual statement that in the process of the multiple production the
information entropy achieves its maximal value for $k=1$ and as a result
$\bar n$ achieves its maximal value and thus does not depend on the
energy at all.  This statement is derived from the fact that this
entropy for NBD near $k=1$ behaves as $\ln \bar n+1-\left({\pi^2
\over 6}-1\right)\left(k-1\right)^2$.  But the lower bound on $k\ge a$
in our model shows that such a statement is a result of the unphysical
interpolation of $k$  to unity.
\bigskip

\noindent {\bf 4. Comparison with experiments}
\medskip

To obtain numerical values of constants $a$ and $b$ in (5) we used
the results of the fit of parameters $\bar n$ and $k$ of NBD by
experimental distributions of charged particles multiplicity in the
range from $\sqrt{s}$ = 19.5 GeV to $\sqrt{s}$ = 900 GeV [10,14,19]
(non-single diffractive events). The result of the fit of $a$ and $b$ in (5)
on the basis of these data gives:
$$a = 3.06 \pm 0.06, \quad b = 6.95 \pm 0.08. \eqno(18)$$

In Fig. 1 is shown the function (5) obtained with these values for  $a$
and $b$.  We did not consider points corresponding to low energy (see
footnote 2).  Fig. 2 shows the dependence of $1/k$ on $\bar n$, which is
seen to saturate.  Fig. 3 gives the curves for $C_q(q=2-5)$ for our
model (solid line) and compares them with experimental data [10,14,19]
for non-single-diffractive component of $pp(\bar pp)$ reactions.
It is seen that higher moments $(q=4,5)$ have not yet achieved their
asymptotic values ($C_q$ = 4.39 and $C_5$ = 10.19 at $\sqrt{s}$ = 900 GeV.

In Fig. 4-7 are shown the distributions $\psi(z,k)$  as a function of
$z=n/\bar n$ for energies 0.546, 1.8, 8, and 40 TeV, respectively.
Solid lines show the asymptotic distribution (12).  It is seen from
these figures that scaling sets in at $\sqrt{s} \approx$ 8 TeV.  These
curves show the systematic shift to the left of the peaks of $\psi(z,k)$
with increasing $\sqrt{s}$.

Finally, Fig. 8 shows the information entropy
$$w=-\sum_n P_n\ln P_n \approx \ln\bar n - \int_0^{\infty} \psi(z,k)
\ln \psi(z,k) dz \eqno(19)$$
which is defined by the chaotic component only for $k$ from (5).  The
figure also shows the ``maximal'' entropy $w_{\rm max}$ corresponding to
$k=1$ which is meaningless in our model.
\bigskip

\noindent {\bf 5. Conclusions}
\medskip

In the present paper we have attempted to establish the relation
between parameters of $k$ and $\bar n$ of NBD on the basis of the general
theory of random branching processes.  This relation seems to be
rather interesting, selfconsistent and has a predictive power.  It
removes some contradictions which occured in the use of NBD for
description of multiplicity distributions for high energy hadronic
collisions.

On the whole the agreement of our model with the existing experimental
data is good enough which, of course, is not surprising because of the
coefficients $a$ and $b$ in (5) were derived from a fit to the experimental
data for $k$ and $\bar n$.  More important are the predictions for the
behavior of $C_q\left({D_2\over\bar n}\right)$ and $\psi(z,k)$ at higher
energies which may be checked at LHC and SSC: the restoration of the KNO
scaling in the multi-TeV region, asymptotic constant values of $C_q$,
depending on $q$, ``explanation'' of the Wroblewski rule at high energy,
and the asymptotic value of the peak of $\psi(z)$.

It is interesting to compare qualitatively our model of general
branching processes with the detailed models of quark gluonic
branching processes.  If one neglects the quark branching the resulting
parton distribution looks very similar to NBD and their conclusions
qualitatively coincide  with ours (limit of the widening of the
distribution shape, increase and final saturation of $C_q$, etc. [21].)
The dominant role of gluonic branching in comparison with quark branching
is the characteristic feature of the detailed study of corresponding
processes from the point of view of dynamical chaos [22], or from the
approach based on the detailed consideration of branching of quarks and
gluons at the formation of quark-gluon plasma [23,24].

There is at least one aspect which apparently necessitates the quark
branching:  the observed small oscillations in the high-multiplicity
tail of $P_n$-distribution at Tevatron energy [25].  If we recall the
very old prediction of such oscillations in the Regge-pole approach [26]
which is connected with Pomeron cuts, then it seems reasonable that
quarks may be responsible for these phenomena.  (The ``explanation'' of these
oscillations by the addition of two binomial distributions
(five-parameter fit [25]) may also be the reflection of this
two-Reggeon cut.

Finally in connection with the meaning of parameter $k$ of NBD and its
asymptotic limit in our model $(k_{\rm min}\approx 3)$ it would be very
interesting to apply our model to the multiplicity distribution of
hadrons in $\pi p$, as well in $e^+e^-$, $\nu p$ and $ep$ collisions
at high energies.

In conclusion we thank to I. G. Aznaurian and N. L. Ter-Isaakian
for discussions.  One of us (S.G.M) is grateful to Berndt M\"uller for
interesting discussions and useful advice, and to C. Gong and C. Wang
for help with the fit of the experimental data.  S.G.M. thanks the
Department of Physics, Duke University for hospitality.  This work was
supported in part by the Department of Energy (Grant DE-FG05-90ER40592).
\vfill
\eject

\noindent {\bf References}
\medskip

{\frenchspacing
\item{[1]} G. J. Alner et al. {\sl Phys. Lett. {\bf 138B}}, 304 (1984).

\item{[2]} A. Giovannini and L. Van Hove, {\sl Acta Phys. Pol. {\bf
B19}}, 495, 917, 931 (1988).

\item{[3]} A. Giovannini, {\sl Proc. Intern. Conf. on Physics in
Collision}, 6, ed. M. Derrick (World Scientific, Singapore, 1987) p.
39.

\item{[4]} P. V. Chliapnikov and O. G. Tchikilev, {\sl Phys. Lett. {\bf
222B}}, 152 (1989).

\item{[5]} G. Altarelli and G. Parisi, {\sl Nucl. Phys. {\bf B126}}, 298
(1977).

\item{[6]} L. Van Hove and A. Giovannini, {\sl Proc. 25th Intern. Conf.
High Energy Physics}, V. II Singapore p. 998, (1990).

\item{[7]} P. Carruthers and C.-C. Shih, {\sl Phys. Lett. {\bf 137B}},
425 (1989).

\item{[8]} G. N. Fowler et al., {\sl Phys. Rev. Lett. {\bf 56}}, 14
(1986).

\item{[9]} P. A. Carruthers et al. {\sl Phys. Lett {\bf B309}}, 369 (1988).

\item{[10]} R. E. Ansorge, {\sl Z. Phys. {\bf C43}}, 357 (1989).

\item{[11]} S. G. Matinyan, {\sl Sov. J. Part. Nucl. {\bf 16}}, 226 (1985).

\item{[12]} B. A. Bambah and M. Venkata Satyanarayana, {\sl Phys. Rev.
{\bf D37}}, 2202 (1988).

\item{[13]} A. Vourdas and R. M. Weiner, {\sl Phys. Rev. {\bf D38}}, 2209
(1988).

\item{[14]} G. J. Alner et al., {\sl Phys. Lett. {\bf 167B}}, 476 (1986).

\item{[15]} V. Gupta and N. Sarma, {\sl Z. Phys. {\bf C41}}, 415 (1988).

\item{[16]} A. K. Chakrabarti, {\sl Phys. Rev. {\bf D45}}, 4057 (1992).

\item{[17]} N. A. Dmitriev and A. N. Kolmogorov, {\sl Dokl. Acad. Nauk. SSSR,
{\bf 56}}, 7 (1947) (in Russian).

\item{[18]} R. Schwed, G. Wrochna and A. K. Wroblewski {\sl Acta Phys. Pol.
{\bf 19B}}, 703 (1988).

\item{[19]} G. J. Alner et al., {\sl Phys. Rep. {\bf 154}}, 247
(1987).

\item{[20]} E. G. Boos et al., {\sl Proc. 25th Intern. Conf. High
Energy Physics}, Singapore, Vol. II, p. 1018 (1990).

\item{[21]} I. Sarcevic, {\sl Mod  Phys. Lett. {\bf A2}}, 513 (1987).

\item{[22]} B. M\"uller, Duke University preprint, Duke-TH-92-36.

\item{[23]} K. Geiger and B. M\"uller, {\sl Nucl. Phys. {\bf B369}}, 600
(1992).

\item{[24]} E. Shuryak, {\sl Phys. Rev. Lett. {\bf 68}}, 3270 (1992).

\item{[25]} C. S. Lindsey, {\sl Nucl. Phys. {\bf A 544}}, 343 (1992).

\item{[26]} V. A. Abramovski and O. V. Kancheli, {\sl JETP Lett. {\bf
15}}, 397 (1972); V. A. Abramovski, V. N. Gribov and O. V. Kancheli,
{\sl Sov. J. Nucl. Phys. {\bf 18}}, 308 (1979).

\item{[27]} G. I. Alner et al., {\sl Phys. Lett. {\bf 160B}}, 199
(1985).
\vfill
\eject

\noindent {\bf Figure Captions:}
\bigskip

\itemitem{Fig. 1:} The dependence of $k$ on $\bar n$ from eq. (5)
with coefficients $a=3.06\; b=6.95$ obtained by fit.  Two points
shown on Fig. 1 and corresponding to low energies ($\sqrt{s} < 19.5$
GeV), are not taken into account (see footnote 2).
\bigskip

\itemitem{Fig. 2:} ${1\over k}$ dependence on $\bar n$ from (5) ($a=
3.06,\; b=6.95$).
\bigskip

\itemitem{Fig. 3:} The $C_q$-moments $(q=2-5)$ as a function of $\bar
n$ from (5) (solid lines) compared with experimental data on inelastic,
non-single-diffractive component of $pp(\bar pp)$ reactions (see table
2 from [27] and [10,14]).
\bigskip

\itemitem{Fig. 4:} The dependence of $\bar n P_n$ on $z={n\over\bar
n}$.  Dashed line for $\sqrt{s}$ = 546 GeV, solid line is an
asymptotic distribution.
\bigskip

\itemitem{Fig. 5:} Same for $\sqrt{s}$ = 1800 GeV.
\bigskip

\itemitem{Fig. 6:} Same for $\sqrt{s}$ = 8 TeV.
\bigskip

\itemitem{Fig. 7:} Same for $\sqrt{s}$ = 40 TeV.
\bigskip

\itemitem{Fig. 8:} The information entropy as a function of $\bar n$.
Solid line is for our model with $k$ dependence of (5).  Dashed-dotted
line corresponds to the ``maximal'' entropy $(k=1)$.

\end